\documentclass[manuscript]{acmart}

\usepackage{booktabs}
\usepackage{graphicx}
\usepackage{xcolor}
\usepackage{tikz}
\usetikzlibrary{arrows.meta,positioning,calc}
\usepackage{array}
\usepackage{caption}
\usepackage{enumitem}
\usepackage{hyperref}
\definecolor{CAnth}{HTML}{0072B2}
\definecolor{CDef}{HTML}{E69F00}
\definecolor{CNeg}{HTML}{D55E00}
\definecolor{CPos}{HTML}{009E73}
\definecolor{CSame}{HTML}{7A7A7A}
\definecolor{CSky}{HTML}{56B4E9}
\definecolor{CInk}{HTML}{222222}
\definecolor{CMute}{HTML}{5C5C5C}
\definecolor{CLine}{HTML}{D8D8D8}
\definecolor{CBg}{HTML}{F3F6F9}
\definecolor{CBg2}{HTML}{FFF8EC}
\definecolor{CHead}{HTML}{1B4F72}
\definecolor{CPaper}{HTML}{FAFBFC}

\colorlet{ink}{CInk}
\colorlet{muted}{CMute}
\colorlet{claim}{CAnth}
\colorlet{neg}{CNeg}
\colorlet{bar}{CAnth}
\colorlet{graybar}{CDef}
\colorlet{ok}{CAnth}
\colorlet{paperbg}{CPaper}

\setcopyright{rightsretained}
\acmConference[CHI '27]{CHI Conference on Human Factors in Computing Systems}{May 10--14, 2027}{Pittsburgh, PA, USA}
\acmYear{2027}


\title[Anthropomorphic Attitudes Divide User Behavior]{How Anthropomorphic Attitudes Toward Agents Divide User Behavior, and How Design Should Change}

\author{Tingyu Cao}
\affiliation{%
  \institution{Tsinghua University}
  \city{Shenzhen}
  \country{China}}
\email{cao-ty25@mails.tsinghua.edu.cn}

\author{Yan Tang}
\affiliation{%
  \institution{Tsinghua University}
  \city{Shenzhen}
  \country{China}}

\author{Keer Hu}
\affiliation{%
  \institution{Tiantong Storm Technology Co. Ltd.}
  \city{Shenzhen}
  \country{China}}

\author{Bomingxuan Zhu}
\affiliation{%
  \institution{Shenzhen MSU-BIT University}
  \city{Shenzhen}
  \country{China}}

\author{Ruyi Zou}
\affiliation{%
  \institution{Shenzhen University}
  \city{Shenzhen}
  \country{China}}

\author{Huaze Tang}
\affiliation{%
  \institution{Tsinghua University}
  \city{Shenzhen}
  \country{China}}
\email{thz21@mails.tsinghua.edu.cn}

\author{Yuanbo Tang}
\affiliation{%
  \institution{Tsinghua University}
  \city{Shenzhen}
  \country{China}}
\email{tyb22@mails.tsinghua.edu.cn}

\author{Shao-Lun Huang}
\authornote{Corresponding authors: Shao-Lun Huang (shaolun.huang@sz.tsinghua.edu.cn) and Yang Li (yangl@cuhk.edu.cn).}
\affiliation{%
  \institution{Tsinghua University}
  \city{Shenzhen}
  \country{China}}
\email{shaolun.huang@sz.tsinghua.edu.cn}

\author{Yang Li}
\authornote{Corresponding author.}
\affiliation{%
  \institution{The Chinese University of Hong Kong (Shenzhen)}
  \city{Shenzhen}
  \country{China}}
\email{yangl@cuhk.edu.cn}

\begin{document}

\begin{abstract}
Studies of conversational agents rely on interviews, surveys, lab tasks, or taxonomies. Whether anthropomorphic naming splits those patterns, and whether replies should change, lack field evidence. We report a field study of a desktop assistant (May--August 2026; $n=257$). Stance is an anthropomorphic assistant name (118 users; 45.9\%). Behavior comes from a 128-unit sparse-autoencoder codebook, identity tags, and opening lexicons. Ranks differ by stance (Spearman $\rho=+0.249$ vs.\ null mean $+0.382$; $p=.0419$). Teacher and tutoring identities covary with renaming, an introverted task-only orientation against it; renamers greet more and command less; confiding shows the codebook's largest gap. We revise the agent: greet and receive affect on relational openings, execute directly on tasking. In 1{,}744 judgments ($n=129$), renamers chose the revised reply far more often than default-name users on being treated as a person and having affect received, less robustly on willingness to continue, and neither judged it worse on task interference.
\end{abstract}

\begin{teaserfigure}
\centering
\includegraphics[width=\textwidth,height=0.26\textheight,keepaspectratio]{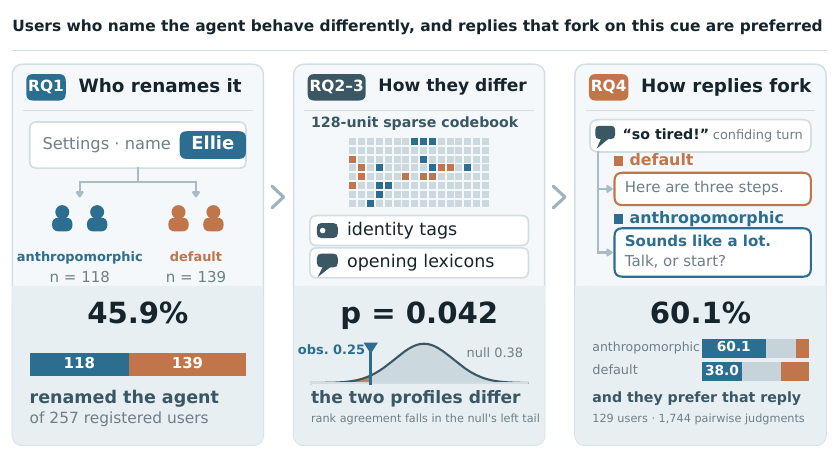}
\caption{Visual abstract. The three columns follow the three stages of the study; within each column, the study design sits above the corresponding result.}
\label{fig:overview}
\Description{Three panels. Left (RQ1): the assistant-name field in settings splits users into an anthropomorphic group of 118 who renamed the agent and a default group of 139 who kept the default name; below, 45.9 percent and a stacked bar of 118 against 139. Middle (RQ2--3): a 128-cell sparse codebook grid plus identity tags and opening lexicons; below, $p=.042$ and a permutation null curve on which the observed value 0.25 falls in the left tail of a null centered at 0.38. Right (RQ4): a single confiding opening turn, ``so tired!'', forks into two replies: the default branch answers the task in one move, while the anthropomorphic branch first receives the affect and then asks whether the user would rather talk or start; below, 60.1 percent and segmented bars showing that the anthropomorphic group chooses the anthropomorphic reply on 60.1 percent of judgments against 38.0 percent in the default group.}
\end{teaserfigure}

\enlargethispage{2\baselineskip}
\maketitle

\section{Introduction}
\label{sec:intro}

Conversational agents have become everyday tools. What people leave behind is not a short laboratory task but a long, voluntary record of real behavior in real settings. That record already supports population-level description: ChatGPT messages divide into asking, doing, and self-expression~\cite{chatterji2025}; long-running donated chats can be labeled advisor, assistant, or companion~\cite{karnam2026}. Human--agent interaction logs contain observable regularities in human behavioral patterns.

In the search for such patterns, anthropomorphism offers a key lead. Early work established that people carry social responses over to computers, responding to media much as they respond to other people~\cite{nass1994,reeves1996}, and naming a device has long been read as a visible trace of that relationship, as with Roomba owners who name the vacuum~\cite{sung2007}, Alexa reviewers who address it by a person-name and treat it as a person~\cite{purington2017}, and older adults who hesitate between calling the assistant ``she'' and ``it''~\cite{pradhan2019}. More recent work shows that anthropomorphic language is itself something that can be identified and scored~\cite{devrio2025,cheng2024}. Product-side research, however, has largely pursued a single direction, asking how users respond when the system is designed to seem more human~\cite{araujo2018,haugeland2022,skjuve2021,skjuve2022}. The complementary side remains underexplored: among users who themselves choose to treat the system anthropomorphically, do real behavioral patterns differ systematically, and can that difference in turn be used to improve the system?

These two lines of work have not been joined. Usage studies can reduce long-running logs to labels such as asking, doing, or companion, but the labels stop at description: knowing what a population does does not say how the model should answer the next turn~\cite{vandermaden2026}. Anthropomorphism studies, in turn, mostly change how human-like the system is, rather than conditioning the reply on a stance the user has already taken. Product changes therefore remain stuck at two poles: one register for everyone, which cannot cover already-split usage, or a per-person profile that is costly and privacy-heavy~\cite{kirk2024}. This paper asks whether, in real long-term use, people who treat the system anthropomorphically differ systematically in behavioral patterns; and if they do, whether that difference can serve as a condition for improving the reply.

We conduct a field study on a desktop assistant. The analysis window runs from 12~May to 12~August 2026 and covers all 257 registered users in that window, comprising 798 sessions, 4{,}943 user questions, and 5{,}422 assistant replies. The assistant name is a settings field that users may change, and in practice only two behaviors occur: replacing the default with a non-default name, or keeping the default. We therefore operationalize anthropomorphic stance as a binary variable, treating every non-default name as anthropomorphic naming. This operationalization relies only on behavior the product already records; it requires no questionnaire and no per-user profile. Behavioral patterns are then measured by three instruments that share no labels: the name field itself; an $\ell_1$ sparse autoencoder trained on user-question embeddings~\cite{bricken2023,marks2024,cunningham2024sae}; and lexicon coding of opening turns, together with identity tags the system has already written back into settings~\cite{chin2024,devrio2025,karnam2026}.

We ask four research questions in sequence (\autoref{fig:overview}):

\begin{itemize}[leftmargin=1.8em,itemsep=2.5pt]
  \item[\textbf{RQ1}] How do users refer to a general-purpose assistant, and how is anthropomorphic naming distributed in a real population?
  \item[\textbf{RQ2}] Do users who give an anthropomorphic name and users who keep the default differ systematically in their behavioral patterns in real use?
  \item[\textbf{RQ3}] Which user attributes are associated with anthropomorphic naming, and how do the two groups differ in what they do on an opening turn?
  \item[\textbf{RQ4}] Once the reply policy is adjusted to that difference, do users perceive an improvement without a loss in task efficiency?
\end{itemize}

RQ1 and RQ2 follow a question form established in the anthropomorphism literature. Purington et al.\ coded Amazon Echo reviews for whether the writer used the name \emph{Alexa} or a personal pronoun, treated the result as a continuum of personification, and tested how it related to the sociability of interactions and to user satisfaction~\cite{purington2017}; Pradhan et al.\ examined, through interviews and usage logs, how older adults move between calling the assistant ``she'' and ``it''~\cite{pradhan2019}. We move the same question into a long-running deployment of a general-purpose assistant. The form of address no longer has to be coded out of review text: it is a field the user has already written into settings. And what it is related to is no longer self-reported satisfaction, but three months of conversational behavior that actually took place.

The paper makes three contributions. First, it proposes a lightweight, observable operationalization of anthropomorphic stance: one naming act the user has already performed separates treating the assistant as a conversational partner from treating it as a tool, and we show that this distribution is far from uniform in a real population. Second, it provides empirical evidence that the two groups' behavioral patterns differ systematically, and further locates the difference: which user attributes covary with naming, and what those users do when they open a turn. Third, it designs and evaluates in the field a differentiated reply policy conditioned on stance and on the current opening type, rather than giving all users the same new persona.

\section{Related Work}
\label{sec:related}

\subsection{Anthropomorphism: a design view and a user-stance view}
The media equation and the CASA paradigm established that people respond socially to computers~\cite{reeves1996,nass1994}, and the three-factor account explains when anthropomorphism occurs through elicited agent knowledge, effectance, and sociality motivation~\cite{epley2007}. Most work on conversational agents and large language models takes a \emph{design view}, asking how anthropomorphic cues affect trust, self-disclosure, and reliance~\cite{araujo2018,abercrombie2023,ischen2020,pentina2023,humanizing2025,cohn2024}. Studies of real logs show that asking and doing remain the majority while companionship is a minority~\cite{chatterji2025,karnam2026}, and that who anthropomorphizes is uneven within a population~\cite{chin2024,skjuve2021,skjuve2022,manoli2025}. Linguistic expressions that make a system sound human can be taxonomized~\cite{devrio2025} and scored~\cite{cheng2024}. This paper takes the complementary \emph{user-stance view}: we first read the stance the user has already brought to the interaction, and only then ask how design should respond.

Naming has long been treated as a marker of social relationship~\cite{purington2017,lopatovska2018,sung2007,pradhan2019}. Those studies read names and pronouns in reviews or interviews; we read a name already written in settings. Every non-default name counts as anthropomorphic naming, with no further tiers. Because naming almost always occurs at first contact, it is a stable stratifying variable rather than a state variable that updates as the conversation proceeds.

The distinction between the two views matters for what a result can be used for. Under the design view, the manipulated quantity belongs to the system, so a finding translates directly into a build decision: add a name, add a face, soften the register. Under the stance view the quantity belongs to the user, and the corresponding decision is not what to build but whom to build it for. This is why the operationalization must stay cheap. A disposition scale would give a finer measurement of stance but cannot be administered to a live population at every session, and a learned per-user profile reintroduces exactly the cost and privacy burden that motivated the question~\cite{kirk2024}. A field that the user has already filled in is the weakest instrument that still separates the two populations, and \S\ref{sec:rq1} shows that it separates them into groups of comparable size.

\subsection{Methodological approaches to studying agent use}
HCI has long studied how people use conversational agents through three families of methods, each with its own cost. The first is interviews and questionnaires. Skjuve et al.\ characterized through interviews how relationships with chatbots form and develop over time~\cite{skjuve2021,skjuve2022}, and Chin et al.\ measured anthropomorphic tendency with a scale and related it to word use~\cite{chin2024}. These methods reach motives and self-understanding, but they are subject to recall and self-report bias and are hard to sustain over long, low-frequency real use. The second is laboratory tasks and controlled deployments; Haugeland et al., for instance, compared topic-guidance strategies under controlled conditions~\cite{haugeland2022}. Such designs support causal claims, but the task is supplied by the researcher, which removes the very dimension of why the user came. The third is log analysis: Chatterji et al.\ estimated the proportions of asking, doing, and self-expression over large volumes of ChatGPT messages~\cite{chatterji2025}, and Karnam et al.\ annotated donated long-term conversations for advisor, assistant, and companion relations~\cite{karnam2026}. Logs preserve the real setting and its duration, at the cost of observing behavior but not intent, and of being almost impossible to randomize.

This paper belongs to the third family, with two adjustments. First, the grouping variable comes neither from a questionnaire nor from post-hoc coding, but from a choice the user makes in settings, in almost every case at first contact and so before the conversations we analyze; it therefore precedes the behavior we want to explain and is untouched by researcher intervention. Second, we do not stop at describing a population: RQ4 takes the difference found in the logs back to the system and tests, through in-product pairwise judgments, whether the resulting change is perceived as an improvement. This path, from field logs to a design intervention and back to field evaluation, is precisely the step that van der Maden et al.\ identify as missing from most current usage analysis~\cite{vandermaden2026}.

\subsection{Representing behavioral patterns: fixed taxonomies and sparse codebooks}
We define a behavioral pattern as shared, sparse, and composable content or practice: one turn may both dispatch a task and offer a backchannel. Finding such patterns in logs asks two things of the representation: it must be composable, so that one turn can carry several components at once, and it must be open, so that the categories come from the data rather than being fixed in advance. The two representations that log studies commonly adopt each satisfy only one of these. A single label per user (advisor, assistant, or companion~\cite{karnam2026}) is not composable, so it cannot express a turn that dispatches a task and maintains a relationship at once. A fixed taxonomy of intents~\cite{chatterji2025} is composable but not open: it can only find the practices its category list already contains, which is a problem when the object of interest is precisely what the designers of the taxonomy did not anticipate. A sparse codebook satisfies both requirements, at the cost of requiring a separate interpretation step before any unit can be discussed.

Sparse autoencoders estimate interpretable codebooks over representations~\cite{bricken2023,templeton2024,cunningham2024sae}, and reading top activations is itself a separate interpretation task~\cite{marks2024}. Recent methodological work has also clarified the limits of such decompositions. Chanin et al.\ show that when the underlying concepts form a hierarchy, the sparsity objective induces \emph{feature absorption}: a latent that appears monosemantic stops firing on part of its domain because its direction has been taken over by a more specific child latent, and neither varying dictionary size nor varying sparsity removes the problem~\cite{chanin2025absorption}. Bussmann et al.\ frame the choice of dictionary size as a genuine tension: a larger dictionary covers more concepts but makes splitting and absorption more likely~\cite{bussmann2025matryoshka}. In a systematic evaluation, Karvonen et al.\ find that gains on unsupervised proxy metrics do not reliably transfer to downstream tasks~\cite{karvonen2025saebench}. Features learned on the same data under different seeds need not align, and a ReLU$+\ell_1$ objective is more stable than TopK~\cite{paulo2026sae}.

Together these results define how we use the codebook: we freeze the dictionary at a single seed, treat the codebook as a \emph{useful decomposition} rather than the unique atomic representation of dialogue, make no claim that individual latents are monosemantic, and place the inference on the whole profile rather than on any single latent (\S\ref{sec:stats}). Per-latent $p$-values are corrected with the Benjamini--Hochberg procedure~\cite{bh1995}; we do not claim equivalence to knockoff selection~\cite{candes2018knockoffs,enkhbayar2025knockoffs}.

When writing short names for latents we follow the \emph{closing procedure} of Li et al.~\cite{li2020np}: open coding of real turns, convergence to a finite category table, and double coding. Our categories are not their non-progress types; we label whether a latent, when it fires, corresponds to holding a relationship, dispatching a task, running a procedure, or talking about the system itself.

\subsection{What an adaptive reply should be conditioned on}
Personalized dialogue typically conditions on a stable persona~\cite{zhang2018persona,li2016persona}, and empathic generation typically conditions on detected affect~\cite{rashkin2019,ma2020}. Chin et al.\ recommend offering more support to highly anthropomorphizing users~\cite{chin2024}. We keep that direction but push the conditioning variable down to the \emph{current opening type}: greeting, confiding, or tasking. The reason appears in \S\ref{sec:rq3}: user attributes indicate who is more likely to rename, but they do not license switching persona by occupation, whereas the difference in opening behavior appears exactly where a runtime system can read it.

The two conditioning choices also fail differently, which is what makes the distinction practical rather than terminological. A persona conditioned on the person is wrong for a whole population whenever the stratification is wrong, and the user has no way to correct it within a session. A policy conditioned on the turn is wrong only for the turn, and the next turn re-reads the condition. In a deployed assistant where stance is inferred rather than declared, that difference in blast radius is the main reason we attach the switch to the opening rather than to the profile. It is also why \S\ref{sec:rq4} leaves the tasking path unchanged between branches and reserves one evaluation item for interference: a turn-level policy is only safe if the path that receives no affective response is left genuinely unchanged.

\section{Method}
\label{sec:method}

\subsection{Setting and data}
\label{sec:data}
The setting is a desktop assistant in which users can rename the assistant in settings. We made no interface changes during log collection. The analysis window is \textbf{12~May to 12~August 2026}, and the study population is all 257 registered users in that window. All records were de-identified before analysis (\S\ref{sec:ethics}). User-chosen assistant names appear throughout as pseudonyms of the same type, so a name that reads as a person is reported as one. Individual users quoted below are labeled U1--U10 in order of first appearance, and the labels are stable across sections so that repeated mentions of the same account can be recognized.

Conversation records are the primary material: \textbf{798} sessions, \textbf{4{,}943} user questions, \textbf{5{,}422} assistant replies, and \textbf{15{,}066} internal tool calls, for 25{,}431 record lines in total. Counting internal steps (thinking, tool returns, and intermediate output), the total is 39{,}191 steps. Usage is highly uneven across users. The product also offers proactive services such as morning briefings, reminders, weekly and monthly reviews, and subscription alerts, which show that the assistant can initiate a conversation; these describe the completeness of the setting and are not part of the present analysis.

Three opening types already occur naturally in the raw logs, and they later form the fork condition in \S\ref{sec:rq4}:

\begin{itemize}[leftmargin=1.1em,itemsep=2pt]
  \item Greeting. U1: ``Ellie, hello, this is the first time we have met.''
  \item Confiding. U2: ``I've given up on it, I feel awful, my enthusiasm is worn down.''
  \item Tasking. U3: ``Just complete it, stop asking me questions.''
\end{itemize}

\subsection{Operationalizing anthropomorphic stance}
\label{sec:stance}
The only grouping variable is the assistant-name field in the user table. Users who replaced the default form the anthropomorphic group ($n=118$); users who kept the default form the default-name group ($n=139$). Every non-default name counts as anthropomorphic naming. We do not measure disposition with a questionnaire, and we do not infer group membership from a user profile.

\subsection{Three instruments for behavioral patterns}
\label{sec:instruments}

\paragraph{Instrument 1: the name field.} The settings-layer assistant name is both the object described in RQ1 and the dependent or grouping variable in RQ2 and RQ3.

\paragraph{Instrument 2: sparse autoencoder.} User questions are embedded with Qwen \texttt{text-embedding-v4} ($d=1024$), and an $\ell_1$ sparse autoencoder is trained on the deduplicated question set: the questions of \S\ref{sec:data} with repeated records of the same question collapsed to their first occurrence.
The encoder is $z=\mathrm{ReLU}(W_{\mathrm{enc}}(x-b_{\mathrm{dec}})+b_{\mathrm{enc}})$ and the decoder is $\hat{x}=W_{\mathrm{dec}}z+b_{\mathrm{dec}}$, with decoder columns unit-normalized after each update. The loss is
\[
\mathcal{L}=\|x-\hat{x}\|_2^2+\lambda\|z\|_1 ,
\]
implemented as batch-mean MSE plus $\lambda$ times the batch-mean $\ell_1$ term, with $\lambda=3\times10^{-3}$. We use Adam, learning rate $10^{-3}$, batch size 512, 400 epochs, and seed~7. The hidden width is $m=256$, and the codebook retains the $K=128$ latents with the highest activation frequency. Each user's behavioral profile is the vector of per-latent activation rates over those 128 latents. Short names for latents are written independently from the five most strongly activating questions, without access to group labels or any statistic~\cite{marks2024,cunningham2024sae}; the temperature is fixed at 0 and a second prompt cross-checks each label.

\paragraph{Instrument 3: identity tags and opening lexicons.} The overall design follows Chin et al.~\cite{chin2024}: characterize who the user is on one side, and count the use of particular words on the other. We change one element: we do not administer a questionnaire. The system already writes a user profile back into settings, and we read those identity tags directly (teacher, tutoring, employee, professional-exam preparation, and others). Opening behaviors are coded with lexicons: a model first inspects roughly 10\% of the data to induce candidate opening practices, and the resulting lexicon is then applied to the full set. It covers opening greeting (\emph{hello} / \emph{hi} / \emph{good morning} / \emph{are you there} within the first five turns), greeting-prone, first-turn greeting, conversation-only turns (no attachment and no tool call), imperatives (\emph{just do} / \emph{do this for me} / \emph{right now}), politeness, sentence-final softeners, backchannel-prone (a turn consisting only of a short acknowledgement such as \emph{mm-hm} or \emph{ok}), and introverted task orientation. These codes never enter sparse-autoencoder training.

\subsection{Statistical analysis}
\label{sec:stats}
RQ1 is descriptive. RQ2 tests the two group-mean behavioral profiles; the statistic is Spearman $\rho$ between the two rank vectors, with a null distribution constructed by label permutation as detailed in \S\ref{sec:rq2}. In RQ3, binary variables are tested against naming with Yates-corrected $\chi^2$ or Fisher's exact test and continuous scores with Spearman correlation. Benjamini--Hochberg correction is applied within each block as a check~\cite{bh1995}, over the eight tests each block contains. All eight opening behaviors clear it. In the identity-tag block, whose eight tests are the seven identity tags and the one lexicon item reported beside them, teacher, introverted task orientation, and tutoring clear it at $q\approx.03$ while employee and professional-exam preparation do not ($q\approx.07$); we mark those two as suggestive wherever they appear. We print the per-item $p$ throughout. The two blocks are not adjusted the same way, and the difference matters for reading the numbers. Each opening behavior is first reduced to a per-user binary indicator, so for the opening behaviors of \S\ref{sec:rq32} the coefficient we report is the $\phi$ of one $2\times2$ table, which for two binary variables is the same quantity as the Spearman $\rho$ we write throughout, and the $p$-value is that same table's test, with no covariate. The two are not interchangeable, however: the Yates-corrected $p$ depends on the behavior's marginal frequency as well as on the coefficient, so two items can order one way on the coefficient and the other way on the $p$. For the identity tags of \S\ref{sec:rq31} the $p$-value additionally adjusts for the user's daily turn count, because whether the system has written a profile tag back at all depends on how much that user has talked. The one lexicon-coded item reported in that block, introverted task orientation, is adjusted the same way for comparability with the tags it sits beside, although it is not itself written back by the system. The coefficient printed for those items is the unadjusted association, so in that block the coefficient and the $p$-value do not come from the same model.

The instruments do not all cover the same users, so each analysis states its own base and we never pool across bases. RQ1 is descriptive over all 257 registered users, because naming is a settings-layer attribute that can be observed without the user having accumulated any turns. RQ2 needs a behavioral profile for each user and therefore runs over those who contributed at least one deduplicated question; every registered user did so in this window, so RQ2 covers the same $n=257$, with the observed 118/139 split between the two naming groups held fixed throughout the permutation. RQ3 is narrower: an attribute or opening rate estimated from one or two turns is dominated by sampling noise, so both of its subanalyses are restricted to users with at least three questions in the deduplicated set, which leaves $n=103$ (47 anthropomorphic and 56 default-name users). The same noise does not force that threshold on RQ2, because there the estimate is a group mean over more than a hundred users and the null is rebuilt from those same profiles, so whatever a one-question profile contributes enters both sides of the comparison; RQ3 instead estimates an association user by user, where such a profile attenuates the coefficient rather than cancelling. That threshold is the only deliberate exclusion we apply, and we do not filter further on content or language.

Two choices in this design deserve explanation. First, we test the \emph{whole profile} before testing any individual latent. The alternative, running 128 per-latent group comparisons and reporting whichever survive correction, answers a different question. It asks which specific practices differ, which is only meaningful once we know that anything differs at all, and with a population of this size and highly uneven per-user turn counts it has low power for any single latent. Testing the rank agreement of two group-mean profiles pools evidence across all 128 latents into one statistic, so RQ2 can be settled with one test; per-latent statistics, each a two-sided Mann--Whitney $U$ on the two groups' activation rates and corrected across all 128 latents, then serve to locate the difference rather than to establish it. The latent we discuss as L79 (emotional confiding, \S\ref{sec:rq2}) accordingly appears as the largest identifiable gap even though its corrected value does not clear the threshold; we print its uncorrected $p$ and say so at that point.

Second, the null distribution must be built by permuting labels rather than by resampling users or by comparing against $\rho=0$. All users are scored on the same codebook, chosen on the pooled data, so two arbitrary halves of the population already agree substantially in rank; the quantity of interest is the agreement \emph{relative to} that shared baseline. Permuting the stance label while holding every profile $P_i$ and both group sizes fixed reproduces exactly the baseline agreement induced by the shared codebook and by the group-size imbalance, and isolates the contribution of the stance variable. The resulting test is one-tailed on the left, because the substantive hypothesis is that the true grouping is \emph{less} internally consistent across groups than a random grouping of the same people.

\subsection{Ethics}
\label{sec:ethics}
The material analyzed here was produced by users in the ordinary course of using the product, not by a recruited study sample, and the privacy policy accepted at registration provides for research use of de-identified, aggregated data. Before any analysis the corpus was passed through an automated privacy filter: a language model labeled personal and identifying information across all records, and each labeled span was replaced with a stable placeholder, so neither the analysis nor any model trained here had access to identifying content. About five percent of the filtered records were then re-read by hand, independently by two of the authors, and no residual identifying content was found. The sparse autoencoder is a measurement instrument rather than a part of the deployed system: it is fitted once for analysis and never enters the request path (\S\ref{sec:rq41}). The deployment's compliance review confirmed this use. The research provision in the privacy policy governs how the records may be analyzed; what leaves the company in this paper is aggregate, and the quoted utterances were cleared one by one, so the published material contains no personal data of its own. No separate ethics board review is required for secondary analysis of operational logs in this environment; because such requirements vary, we state the safeguards here so that they can be judged directly.

No interface change was made for research purposes during the log window, so no user was assigned to a condition; the naming stance we analyze is a pre-existing feature that users act on for their own reasons. The judgment prompt belongs to the product's own evaluation surface, is voluntary, and carried no reward or consequence. No contact detail or payment information was extracted, and no local file content was read. The identity tags of \S\ref{sec:rq31} are coarse occupational self-descriptions the assistant had already written into its own settings; we read them as they stood, report them only as categories, and extracted no other demographic attribute. Every result is reported in aggregate, and every utterance quoted in this paper was additionally checked by hand.

The conversation logs cannot be released, since they are records of private use by identifiable account holders and de-identification does not make them safe to publish. The analysis code (codebook fitting, the permutation test, and the association tests) is provided as supplementary material; \S\ref{sec:stats} states the estimation procedure in full, so the results can also be checked against the reported statistics without running the code.

\section{Results}
\label{sec:results}

This section answers the first three research questions in turn: how naming is distributed in the population (\S\ref{sec:rq1}), whether the two groups differ in behavioral patterns (\S\ref{sec:rq2}), and where exactly the difference lies (\S\ref{sec:rq3}). The three form a chain of successively tighter evidence: first that a stratification exists, then that it corresponds to a difference in behavior, and finally that the difference sits somewhere a running system can read. The system change built on these results, and its evaluation, are reported separately in \S\ref{sec:rq4}.

\subsection{The naming distribution is not uniform (RQ1)}
\label{sec:rq1}

If every user addressed the agent the same way, a single reply register for everyone would be defensible. Before asking whether the assistant should be ``more human,'' we first count how people actually name it.

\begin{figure}[t]
\centering
\includegraphics[width=\linewidth]{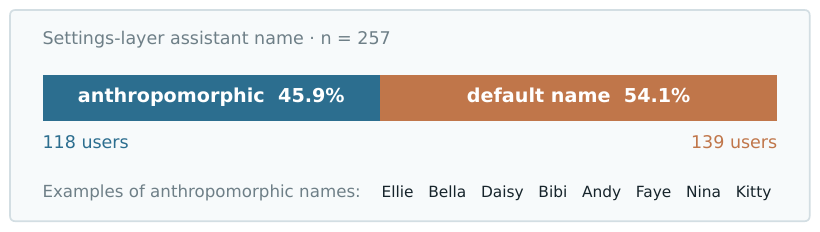}
\caption{RQ1. Settings-layer assistant names across the registered population. Every non-default name counts as anthropomorphic naming.}
\label{fig:naming}
\Description{A horizontal stacked bar: the anthropomorphic segment covers 118 users, 45.9 percent, and the default-name segment covers 139 users, 54.1 percent. Example assistant names appear below the bar.}
\end{figure}

Of the 257 registered users, 118 (45.9\%) set a non-default assistant name and 139 (54.1\%) kept the default (\autoref{fig:naming}). In practice the field admits only these two behaviors. This is a descriptive count and involves no inferential test.

Renamed accounts whose names read as persons, whose users address the assistant as someone present, include U1, who named it \emph{Ellie} and opened with ``Ellie, hello, this is the first time we have met''; U4, who named it \emph{Bella} and opened with ``hey Bella''; U5 (\emph{Daisy}); U6 (\emph{Bibi}); U7 (\emph{Andy}); and U2 (\emph{Faye}).

A single warm persona for everyone is therefore not defensible, since more than half of the population shows no anthropomorphic naming at all. Nor can the name field carry a ``recognize the person, then decide whether to care'' design. The field is readable at request time, and our policy does read it there, but because it is fixed almost at first contact it never reflects anything the system learns about the user afterwards; it can select a branch or serve as a weak prior, and no more. Both points still depend on something RQ1 cannot show: if the two groups did not differ in conversational behavior, a product could simply ignore the field. That is what RQ2 must establish.

\subsection{Behavioral patterns split by naming stance (RQ2)}
\label{sec:rq2}

RQ1 shows only that the settings layer is uneven. To judge whether the product should treat the groups differently, we must show that their conversational behavior actually differs. We compare usage ranks over a shared codebook.

\subsubsection{Unit of comparison and test statistic}
The comparison is between two \emph{group-mean} behavioral profiles, not between each user's single strongest latent. The codebook is selected first, over the deduplicated question set, by activation frequency; this step uses neither user identity nor group label, so both groups share the same 128 coordinates. Each user $i$ receives a full profile $P_i\in\mathbb{R}^{128}$ whose $k$-th entry is the proportion of that user's deduplicated questions on which latent $k$ fires. Writing $\mathrm{A}$ and $\mathrm{D}$ for the anthropomorphic and default-name users within that base ($|\mathrm{A}|=118$, $|\mathrm{D}|=139$), averaging within groups gives $m^{\mathrm{A}}=|\mathrm{A}|^{-1}\sum_{i\in\mathrm{A}}P_i$ and $m^{\mathrm{D}}=|\mathrm{D}|^{-1}\sum_{i\in\mathrm{D}}P_i$, each of which is then ranked from high to low. Spearman correlation between the two rank vectors is $\rho=+0.249$. With no ties,
\[
\rho=1-\frac{6\sum_{k=1}^{K}\left(r^{\mathrm{A}}_k-r^{\mathrm{D}}_k\right)^2}{K(K^2-1)},\qquad K=128 .
\]

\subsubsection{The null is not centered at zero}
The null hypothesis here is not $\rho=0$. Because all users share one codebook, even a random split of the same people into two groups of these sizes yields two rather similar group means. We therefore construct a label-permutation null: each user's profile $P_i$ is held fixed while group labels are shuffled with the observed group sizes preserved, repeated $B=9{,}999$ times with $\rho^*$ recomputed each time. The resulting null distribution has mean about $+0.382$ (sd $0.073$). The left-tail $p$-value is
\[
p=\frac{1+\#\{\rho^*\le\rho_{\mathrm{obs}}\}}{B+1}=.0419 .
\]
Under the true grouping, the two rank vectors are \emph{less} alike than a random cut of the same users would produce (\autoref{fig:sae}A). The test is one-tailed on the left for the reason given in \S\ref{sec:stats}; the corresponding two-sided value is $.084$. This supports the RQ2 conclusion that behavioral patterns split by naming stance.

\subsubsection{Where the difference sits}
\begin{figure}[t]
\centering
\includegraphics[width=\linewidth]{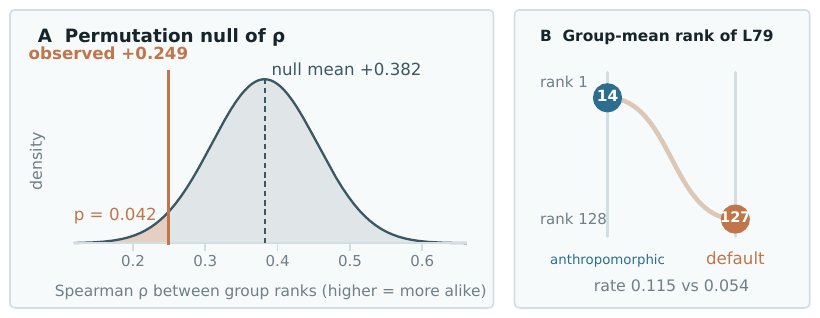}
\caption{RQ2. (A) Distribution of $\rho$ under the label-permutation null; the observed $+0.249$ falls in the left tail, $p=.0419$. The curve is the normal approximation implied by the reported null mean and standard deviation; the permutation null is slightly heavier in the left tail, so the reported $p$ is larger than the area under the drawn curve. (B) Group-mean rank of L79 (emotional confiding): 14th in the anthropomorphic group, 127th in the default-name group. L79 does not survive BH correction and locates the difference rather than establishing it.}
\label{fig:sae}
\Description{Left: a permutation-null density curve with the observed value marked by a vertical line and the left tail shaded. Right: two rank axes joined by a curve, marking rank 14 in the anthropomorphic group and rank 127 in the default-name group.}
\end{figure}

The latent with the largest rank gap is L79 (emotional confiding): group-mean activation rate $0.115$ and rank 14 in the anthropomorphic group, versus $0.054$ and rank 127 in the default-name group (\autoref{fig:sae}B). Its top activations are ``so tired!'', ``fed up lately'', and ``but I'm just sleepy right now.'' These are users reporting their own state to the assistant, not treating emotion as a writing topic. The single-latent $p$ is $.0015$, which does not clear the Benjamini--Hochberg threshold across 128 comparisons. What carries the RQ2 claim is the profile-level permutation test above, which has already rejected; the per-latent screen locates that difference rather than deciding it (\S\ref{sec:stats}). We treat the confiding gap as a substantive result, being both the largest rank gap in the codebook and a more than twofold difference in activation rate, while recording that it is not individually protected against the 128-way multiplicity.

\begin{figure}[t]
\centering
\includegraphics[width=\linewidth]{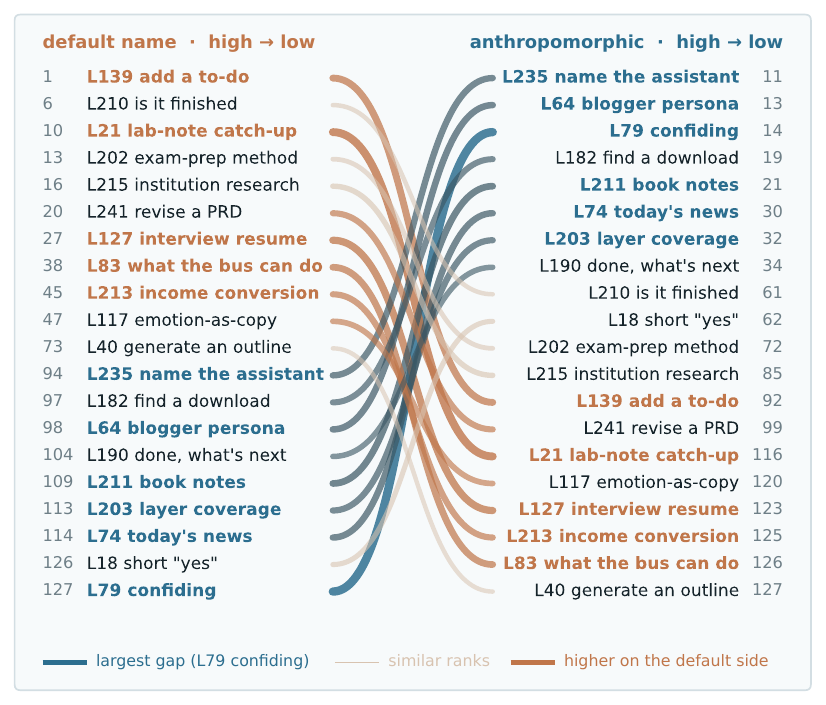}
\caption{RQ2. The 20 labeled behavioral patterns with the largest absolute rank gaps. Left: default-name ranks, high to low. Right: anthropomorphic ranks. Each pattern is joined by a curve whose width grows with the gap. The largest gap is L79, emotional confiding (127th $\rightarrow$ 14th), which does not survive BH correction.}
\label{fig:slope}
\Description{Two-column slope chart. L79 rises from bottom left to upper right; add a to-do and lab-note catch-up fall the other way.}
\end{figure}

The only latent that survives BH correction is L227 (short backchannels such as ``mm-hm'' and ``ok''), ranked 1st versus 26th with $q=.014$. Because it aligns with opening greetings, we discuss it in \S\ref{sec:rq3} rather than as a separate result. On the default-name side, the highest-ranked latents are L139 (add a to-do; rank 1 there against 92 among renamers) and L21 (lab-note catch-up; 10 against 116). \autoref{fig:slope} shows the 20 labeled patterns with the largest absolute rank gaps among the 128 latents.

These results indicate that the two groups differ not only in topics but in practices: the anthropomorphic side shows more relationship-maintaining behavior, and the default-name side more task-dispatch and record-keeping behavior. We stress that naming a latent and testing its rank gap are separate steps: short names are written without access to group labels, so the labels cannot have been produced from the conclusion.

\subsection{User attributes and opening behavior (RQ3)}
\label{sec:rq3}

\begin{figure}[t]
\centering
\includegraphics[width=\linewidth]{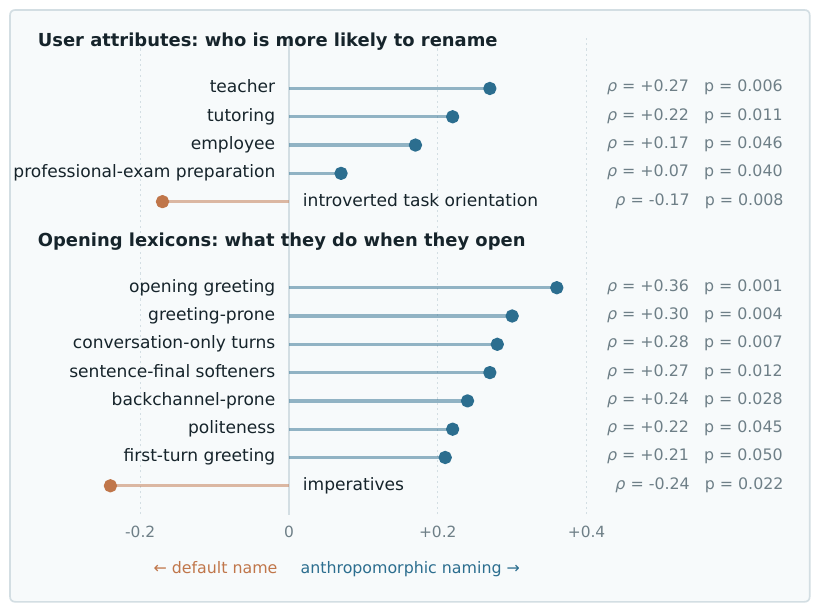}
\caption{RQ3. Associations between anthropomorphic naming and, respectively, user attributes and opening behaviors, over the 103 users with at least three deduplicated questions (47 anthropomorphic, 56 default-name). Points are unadjusted coefficients and every item shown reaches $p\le.05$; the identity-tag $p$-values adjust for daily turn count and the opening-behavior $p$-values do not (\S\ref{sec:stats}). Items extending to the right of zero covary with anthropomorphic naming, those to the left with keeping the default name.}
\label{fig:assoc}
\Description{Horizontal dot plot. The upper block holds four identity tags together with the lexicon-coded introverted task orientation, the lower block eight opening behaviors. Opening greeting has the strongest positive correlation; imperatives and introverted task orientation are negative.}
\end{figure}

RQ2 shows that the two rank distributions differ but does not say where. To turn the result into an implementable design, two finer questions remain: which user attributes covary with anthropomorphic naming, and what the two groups do on an opening turn. Both sets of results are associations (\autoref{fig:assoc}).

Both subanalyses run on the base given in \S\ref{sec:stats}: the 103 users with at least three questions in the deduplicated set, 47 of them anthropomorphic and 56 default-name. As explained there, the opening-behavior figures in \S\ref{sec:rq32} are matched pairs from one test, whereas the identity-tag coefficients in \S\ref{sec:rq31} are unadjusted while their $p$-values adjust for daily turn count, so a small coefficient in that block can carry a smaller $p$-value than the coefficient alone would imply.

\subsubsection{Which user attributes are associated with anthropomorphic naming}
\label{sec:rq31}
User attributes come from the identity tags the system writes back into settings, with introverted task orientation added from the lexicon to measure the execute-only direction; that lexicon item, and not the ``execute-only'' tag itself, is what carries the negative association reported below. On the sparse-autoencoder side, the highest-ranked latents in the default-name group are likewise add a to-do (L139) and lab-note catch-up (L21).

Positively covarying tags are teacher ($\rho=+0.27$, $p=.006$), tutoring ($\rho=+0.22$, $p=.011$), employee ($\rho=+0.17$, $p=.046$), and professional-exam preparation ($\rho=+0.07$, $p=.040$). Introverted task orientation covaries negatively ($\rho=-0.17$, $p=.008$). Student, engineer, and creator tags do not reach significance and are excluded from the conclusion. Of the eight tests in this block, teacher, introverted task orientation, and tutoring clear the Benjamini--Hochberg correction; employee and professional-exam preparation do not, and we carry them as suggestive rather than established (\S\ref{sec:stats}).

Corresponding cases: on the renaming side, U6 and U8 carry a teaching tag, U8 and U2 a professional-exam preparation tag, and U9 and U7 an employee tag. On the other side, U3 (``just complete it, stop asking me questions'') and U10 (``just execute'') carry the execute-only tag. We report the tag categories rather than each user's occupation, because an occupation together with a quoted turn would narrow a user down within a population of this size.

Two caveats apply. First, these effect sizes are generally small, so they support a stratifying judgment about who is more likely to rename, not a per-occupation switch of reply persona. Because the coefficients are unadjusted while the $p$-values adjust for daily turn count (\S\ref{sec:stats}), the two diverge. They diverge most for professional-exam preparation, whose $+0.07$ would not be significant on its own; the point plotted for that item understates its adjusted association by more than any other tag in the block. Second, emotional confiding does not appear in this subsection: it is a turn-level behavior and belongs to \S\ref{sec:rq32}.

\subsubsection{What the two groups do on an opening turn}
\label{sec:rq32}
Opening behavior is coded at the user level by lexicon. Emotional confiding has no lexicon item; it enters through latent L79, whose ranks come from the RQ2 base and are reported alongside these coefficients rather than pooled with them. Positively covarying behaviors are opening greeting ($\rho=+0.36$, $p=.001$), greeting-prone ($\rho=+0.30$, $p=.004$), conversation-only turns ($\rho=+0.28$, $p=.007$), sentence-final softeners ($\rho=+0.27$, $p=.012$), backchannel-prone ($\rho=+0.24$, $p=.028$), politeness ($\rho=+0.22$, $p=.045$), and first-turn greeting ($\rho=+0.21$, $p=.050$). Imperatives covary negatively ($\rho=-0.24$, $p=.022$).

The codebook side agrees. L79 (14th versus 127th) captures emotional confiding, while L227 ($q=.014$; 1st versus 26th) and L18 (a whole turn consisting of ``yes''; 62nd versus 126th) align with greeting. A distinction is needed here: L117 (emotion as a writing topic) and L221 (requesting emotional resonance) are transactional requests and are not merged with L79.

Corresponding cases: greeting, U1 (``Ellie, hello, this is the first time we have met'', ``hi Ellie'') and U2, who repeatedly opens with ``hello''; confiding, U1 (``home is not a harbour for me'') and U2 (``I've given up on it, I feel awful''); tasking, U3 (``just complete it, stop asking me questions'') and U10 (``just execute'').

This subsection yields the most directly implementable result of the paper. Both greeting and confiding occur on the opening turn, which means that waiting until the system has recognized ``what kind of user this is'' comes too late. The conditioning variable available at runtime must be the current opening type: greet when greeted, receive affect when confided in, and execute when given a task.

\section{A Differentiated Reply Policy and Its Field Evaluation (RQ4)}
\label{sec:rq4}

The first three research questions jointly support one design decision: the two groups need not share a single reply policy. More than half of the users keep the default name (RQ1); the two groups' codebook usage ranks differ (RQ2); and the difference can be located in specific user attributes and opening behaviors (RQ3). What we change is therefore the fork condition of the reply policy, not the persona given to everyone.

\subsection{Policy design}
\label{sec:rq41}
The object of the change is the assistant's main system prompt. Previously all users shared one prompt with a uniform task-first register. We keep the hard constraints unchanged (tools must be called, nothing may be guessed, and stated actions must match executed ones) and modify only how an opening turn is answered. This produces two branches. The anthropomorphic branch is the revision. The default-name branch writes the previous task-first register down explicitly, adding no affective move to it, so a user who keeps the default receives what that user would have received before the change. Writing it down does make explicit what was previously implicit, and \S\ref{sec:limit} returns to the fact that we did not run a separate arm to confirm the two are equivalent.

The anthropomorphic branch is enabled when the assistant name is non-default, and splits into three paths by opening type. When the turn is a greeting, the first sentence is a greeting only, followed by asking whether the user would like to talk for a moment or has a specific task; listing to-dos in the first sentence is prohibited. When the turn is confiding, one or two sentences first receive the affect and restate what the user is feeling, without diagnosis or exhortation, and only then ask whether the user prefers to keep talking or move to a task; producing an immediate three-step plan is prohibited. When the turn is tasking, the first item is the deliverable, with no care sentence.

The default-name branch is enabled when the name is still the default and provides no affective response: greetings are not followed by a ``talk or task'' question, confiding is not followed by consolation, and tasking again begins with the deliverable. Both branches share two constraints: the assistant never addresses the user by the assistant's own name, and no occupational persona is added. When a single turn carries a relational move and a task at once (for example, ``so tired, just finish this one for me,'' which confides and assigns in one breath), the tasking path takes precedence.

Opening classification at runtime uses a separate lightweight keyword selector written for the three opening types. It is not the analysis lexicon of \S\ref{sec:instruments}, which has no confiding item, and the sparse autoencoder is not run online either.

\subsection{Evaluation procedure}
\label{sec:rq42}
\begin{figure}[t]
\centering
\includegraphics[width=0.62\linewidth]{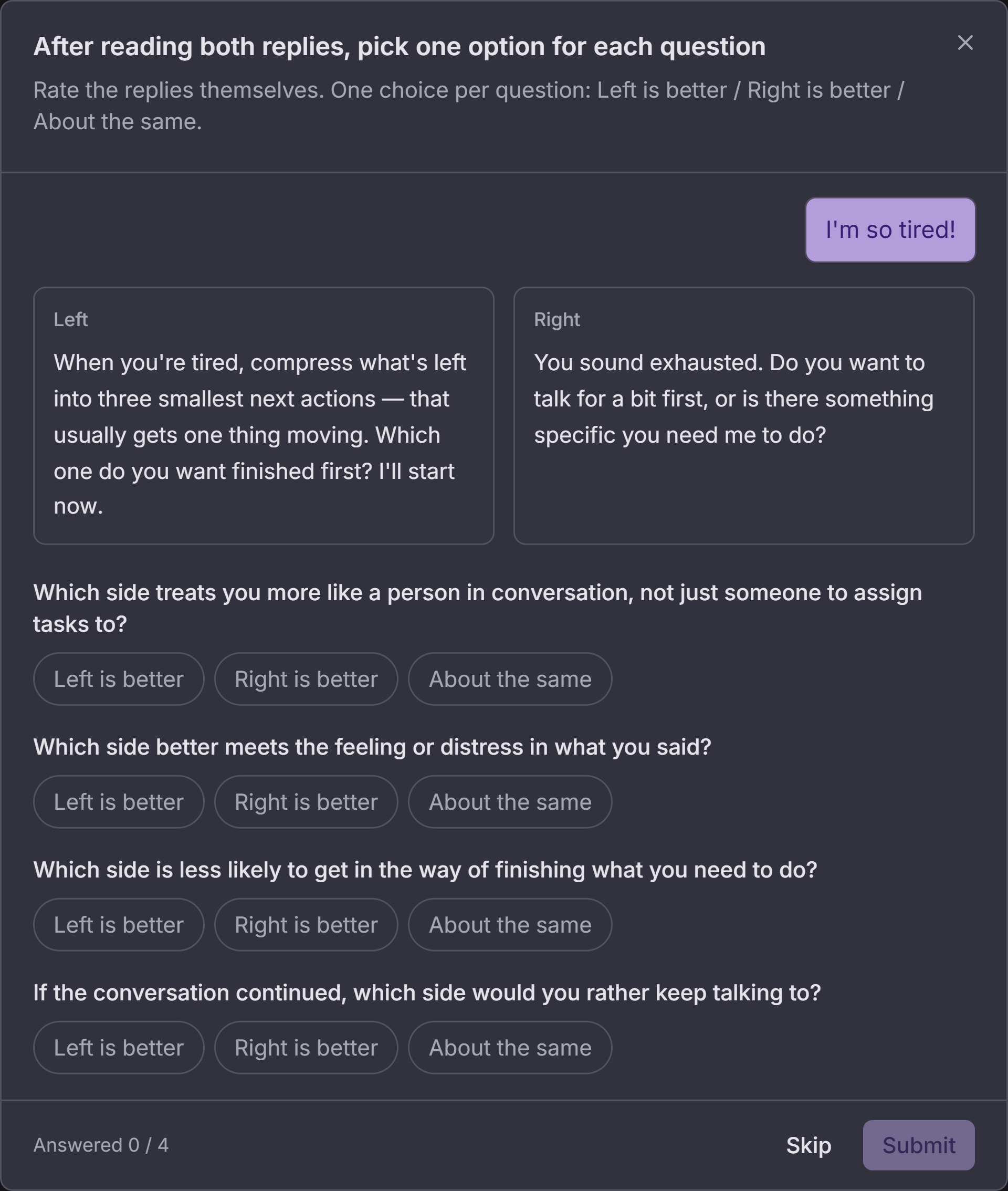}
\caption{The judgment prompt as users saw it, here on the logged opening ``I'm so tired!'' The left reply is the default-name branch and the right the anthropomorphic branch; all four questions are asked at once, and nothing in the prompt names anthropomorphism or indicates which branch produced which reply.}
\label{fig:popupui}
\Description{Screenshot of the in-product dialog. A header reads ``After reading both replies, pick one option for each question.'' Below the user turn ``I'm so tired!'' two candidate replies sit side by side: the left one proposes compressing the remaining work into three small next actions, the right one says the user sounds exhausted and asks whether they would rather talk first or have something done. Four questions follow, each with the buttons ``Left is better,'' ``Right is better,'' and ``About the same,'' above a progress line reading ``Answered 0 / 4'' with Skip and Submit buttons.}
\end{figure}

Each judgment shows two replies, generated side by side~\cite{kahng2024}, to one of the judging user's own logged utterances, which is why the questions can refer to what the user said. The pair is followed by a judgment prompt (\autoref{fig:popupui}). The prompt never mentions anthropomorphism. Each question offers three options: left is better, right is better, or about the same. In the material the left side is always the default-name branch and the right side the anthropomorphic branch. Because the default-name branch reproduces the register in force before the change (\S\ref{sec:rq41}), the left reply is intended to stand for what the same utterance would have drawn from the unmodified assistant, an equivalence we did not test with a separate arm (\S\ref{sec:limit}), and each judgment therefore sets the revision against a written specification of the status quo rather than against a second new alternative. The four questions are:

\begin{enumerate}[leftmargin=1.5em,itemsep=3pt]
  \item Which side treats you more like a person in conversation, not just someone to assign tasks to?
  \item Which side better meets the feeling or distress in what you said?
  \item Which side is less likely to get in the way of finishing what you need to do?
  \item If the conversation continued, which side would you rather keep talking to?
\end{enumerate}

\subsection{Evaluation results}
\label{sec:rq43}
The judgment prompt appears after the two replies are shown side by side and asks all four questions at once, so the number of judgments a user contributes depends on their usage frequency and on the mix of opening types. In total, 129 users took part (59 with anthropomorphic names and 70 with default names), returning \textbf{1{,}744} judgments (962 and 782 respectively). The median number per user is 17 (IQR 9--24) and 11 (IQR 4--20).

\begin{figure}[t]
\centering
\includegraphics[width=\linewidth]{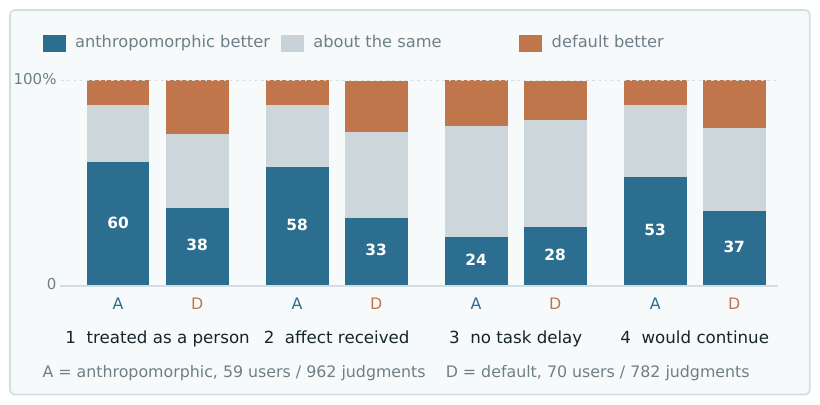}
\caption{RQ4. Choice distributions of the two groups on the four judgment questions. On Q1, Q2, and Q4 the anthropomorphic group chooses the anthropomorphic reply far more often; on Q3 both groups answer mainly ``about the same,'' and neither tilts toward the unmodified reply.}
\label{fig:popup}
\Description{Four pairs of stacked bars, one pair per question. On questions 1, 2, and 4 the anthropomorphic group gives more than half its answers to the anthropomorphic reply; on question 3 the ``about the same'' segment exceeds half in both groups.}
\end{figure}

Results appear in \autoref{tab:popup} and \autoref{fig:popup}. On Q1 (treated as a person in a conversation), the anthropomorphic group chose right 60.1\%, about the same 28.0\%, and left 12.0\%; the default-name group chose 38.0\%, 36.1\%, and 26.0\%. Q2 (receiving affect) points the same way: 58.0 / 30.0 / 12.0 versus 33.1 / 41.9 / 24.9. Q4 (willing to continue): 53.1 / 35.0 / 11.9 versus 36.6 / 40.0 / 23.4. On Q3 (no task delay), both groups answer mainly ``about the same'' (54.0\% and 52.0\%); the remaining choices split evenly among renamers (24.0\% right, 22.0\% left) and tilt slightly toward the right-hand reply among default-name users (28.5\% versus 19.4\%).

For each question we compare the two groups' choice distributions with a $\chi^2$ test of independence on the $2\times3$ table of judgment counts. The groups differ on Q1 ($\chi^2(2)=97.4$, $p<.001$, Cram\'er's $V=0.24$), Q2 ($\chi^2(2)=115.2$, $p<.001$, $V=0.26$), and Q4 ($\chi^2(2)=62.5$, $p<.001$, $V=0.19$), and do not differ on Q3 ($\chi^2(2)=5.1$, $p=.080$, $V=0.05$).

Those 1{,}744 judgments come from 129 people, so a test that treats them as independent observations overstates precision. We bound the consequence by dividing each statistic by the mean number of judgments per user, $1{,}744/129=13.5$, which is the design-effect correction that would apply if everything a user says after their first judgment carried no new information. Under that deliberately extreme discount Q1 ($\chi^2(2)=7.2$, $p=.027$) and Q2 ($\chi^2(2)=8.5$, $p=.014$) remain significant and Q3 remains null ($\chi^2(2)=0.4$, $p=.830$), while Q4 falls to $\chi^2(2)=4.6$, $p=.099$. Intra-user agreement is certainly well below one, so the true values lie between the two calculations. We therefore read Q1 and Q2 as differences that hold under either assumption, Q4 as a difference that is clear at the judgment level but not robust to the worst case, and Q3 as showing no difference at either level. Cram\'er's $V$ is normalized by sample size and is unchanged by the discount.

Three points stand out. First, anthropomorphic users prefer the forked reply on the relational questions, robustly so on Q1 and Q2, which shows that the behavioral difference found in RQ2 and RQ3 corresponds to a perceptible difference in preference. Second, default-name users show a markedly weaker preference on Q1, Q2, and Q4, consistent with the judgment that a single warm persona should not be given to everyone. Third, Q3 gives no sign that the revision interferes with tasks, though what it can support is narrower than it may look. A between-group null would not establish the absence of interference, since two groups can answer alike and both be obstructed; and ``about the same'' compares the two replies with each other rather than saying whether either gets in the way. What the item does show is that the revision is not judged worse than the register it replaced: ``about the same'' is the majority response in both groups (54.0\% and 52.0\%, against 28.0\% and 36.1\% on Q1), and neither group tilts toward the unmodified reply, which takes 212 of the 443 directed judgments among renamers and 152 of 375 among default-name users. Because the two branches occupied fixed sides, we read this as non-inferiority on task interference and not as evidence about efficiency in absolute terms. This downstream test therefore shows that the behavioral difference is perceptible to users and that a policy conditioned on it is preferred by the group it targets; it does not independently validate the codebook statistic, which rests on a different base.

\begin{table}[t]
  \caption{Overall distribution of in-product pairwise judgments (\%). 1{,}744 judgments in total, each covering four questions. Right $=$ the anthropomorphic branch.}
  \label{tab:popup}
  \small
  \setlength{\tabcolsep}{3.5pt}
  \begin{tabular*}{\columnwidth}{@{\extracolsep{\fill}}l*{6}{wc{1.18cm}}@{}}
    \toprule
     & \multicolumn{3}{c}{Anthropomorphic, 59 users $\cdot$ 962} & \multicolumn{3}{c}{Default name, 70 users $\cdot$ 782} \\
    \cmidrule(lr){2-4}\cmidrule(lr){5-7}
    Question & Right & Same & Left & Right & Same & Left \\
    \midrule
    Q1 treated as a person & 60.1 & 28.0 & 12.0 & 38.0 & 36.1 & 26.0 \\
    Q2 affect received & 58.0 & 30.0 & 12.0 & 33.1 & 41.9 & 24.9 \\
    Q3 no task delay & 24.0 & 54.0 & 22.0 & 28.5 & 52.0 & 19.4 \\
    Q4 would continue & 53.1 & 35.0 & 11.9 & 36.6 & 40.0 & 23.4 \\
    \bottomrule
  \end{tabular*}
\end{table}

\section{Discussion}
\label{sec:disc}

\subsection{User stance as a conditioning variable}
The four research questions form one chain. RQ1 shows that the population is uneven in anthropomorphic stance, so a uniform warm persona is not defensible. RQ2 shows that this unevenness extends to conversational behavior: the two groups' rank distributions over a shared codebook differ, and the difference can be located in specific latents. RQ3 further shows that, among the users whose per-user rates can be estimated at all, the difference concerns what they do on opening turns. Because the RQ2 statistic weights every user equally, the users below that threshold contribute to the profile-level result without being represented here, so RQ3 narrows where the difference sits rather than accounting for all of it. That locates it at the place a runtime system can act on, though the coefficients are between-user and do not by themselves establish that any single turn is classifiable; the branch selector is a separate lexicon that classifies the current opening (\S\ref{sec:rq41}). RQ4 designs a fork on that basis and obtains preference evidence inside the product.

Compared with existing adaptive proposals, the difference lies in the choice of conditioning variable. Chin et al.\ recommend more support for highly anthropomorphizing users~\cite{chin2024}; we make ``support'' concrete as three turn-level paths and bound its scope with the distribution from RQ1. Haugeland et al.\ evaluate a version that changes topic guidance for all users~\cite{haugeland2022}; by contrast, we deliberately leave the tasking path unchanged between branches, and Q3 exists to check that the relational moves added on the other two paths do not spill over into it.

\subsection{Implications for HCI research}
Traces users have already left in the interface can serve directly as research variables. Anthropomorphism has mainly been measured with questionnaires~\cite{chin2024} or coded by researchers out of reviews and interview transcripts~\cite{purington2017,pradhan2019}. Both routes require the researcher to intervene after the fact and are therefore exposed to recall bias and to coder judgment. In a deployed product, however, many actions users take for their own purposes (renaming, pinning, setting a reminder) already carry attitudinal information, and they occur before the behavior to be explained. Using such traces as variables costs almost nothing and adds no new privacy burden. The limit is equally clear: a trace records what the user did, not why, so it is suited to stratification and association analysis and cannot substitute for an interview when the question is about motive.

A design goal of ``adapting to the user'' also needs to specify the \emph{unit} of adaptation. Personalization research on dialogue systems has long worked at the level of the person, organized around a stable persona~\cite{zhang2018persona,li2016persona}. Our results point to a finer unit: stance decides \emph{whether} a fork is warranted, and the current opening decides \emph{which branch} to take. The distinction matters because the two fail differently. A misjudgment at the level of the person persists across the whole session and the user cannot correct it from inside the conversation; a misjudgment at the level of the turn costs one turn. For any system in which stance must be inferred rather than declared, that difference in blast radius is what determines whether an adaptive policy can safely ship.

\subsection{Design implications}
For practitioners building a general-purpose assistant, the default register should not be uniformly warm. A recurring product instinct is to make the assistant warmer for everyone, on the assumption that warmth is at worst neutral. RQ1 and RQ4 together argue against that assumption. Slightly more than half of this population never signaled any anthropomorphic stance. Those users did still lean toward the warmer reply on the relational questions, but far less decisively than renamers did: they chose it about a third of the time and the plainer reply roughly a quarter of the time, a margin of roughly three to two, where renamers preferred it by more than four to one. A warmer default is therefore not a free improvement. It buys much less in the larger group than the renaming figures suggest, and on roughly a quarter of that group's judgments the plainer reply was preferred outright. Because judgments are spread unevenly across users, that share is not a count of people, but it does mean a warmer default makes some answers worse.

Our own policy keeps one person-level gate, and it carries the cost that goes with one: the name field decides whether the fork applies at all, so a user who kept the default name and then confides receives no affective response on any turn and cannot change that from inside the conversation. A stance signal and a turn-level signal also do different jobs, and should be read in different places. Because the assistant name is set almost at first contact, it can stratify a population but cannot drive a runtime decision that depends on knowing the user first. The practical consequence is that the useful pairing is a cheap, stable stance variable for \emph{deciding whether a fork exists at all}, and a turn-level signal for \emph{choosing the branch}. In our deployment both are cheap: the stance variable is a settings field, and the branch selector is a keyword lexicon that runs in the request path. Nothing in the runtime policy requires the sparse autoencoder, which is used only in analysis.

Finally, a fork of this kind should be evaluated on interference rather than on preference. It is easy to demonstrate that a warmer reply is perceived as warmer; the substantive risk is that the warmth is paid for in task latency or in extra clarification steps. We left the tasking path unchanged between branches, so that any interference would have to come from the relational moves, and reserved one evaluation item for that question. We would suggest this as a general pattern: when an adaptive policy adds a relational move, at least one evaluation item should be reserved for the population and the path that the move is \emph{not} meant to help.

\subsection{Methodological remarks and transfer}
The sparse autoencoder was chosen because of the nature of the object rather than any claim of higher sensitivity: behavioral patterns are shared, sparse, and composable, which matches the assumptions of a sparse codebook. The most robust signal on the opening turn comes from the greeting lexicon and does not depend on the autoencoder; the codebook contributes a complementary view that presupposes no category list, and it is what makes behaviors such as L79 visible at all. L227 is the only latent surviving BH correction, and we report it as evidence aligned with greeting rather than as a separate finding.

Which parts of this design should transfer, and which should not, follows from where each part is anchored. The procedure is anchored in the product and should transfer: any deployment with a user-editable agent name has a stance variable available at zero measurement cost, and the profile-level permutation test applies to any shared, sparse decomposition of behavior. The specific lexicons are anchored in the conversational conventions of a single language. Sentence-final softeners, for instance, are realized by different devices across languages, so the lexicons would have to be rebuilt for any other language, though the categories they instantiate (greeting, backchannel, imperative, politeness) are not language-specific. The identity tags are anchored in this product's own profile-writing feature and are the least portable element; a deployment without such a feature would need to drop \S\ref{sec:rq31} and rely on RQ2 and \S\ref{sec:rq32} alone, which is sufficient to justify the fork. The effect sizes themselves should not be expected to transfer at all: a desktop assistant used for teaching and tutoring is a particular population, and the same test on a coding assistant or a customer-service agent could plausibly return a null result, which would itself be informative.

\subsection{Limitations and future work}
\label{sec:limit}

The study covers one desktop assistant over a three-month window, with highly uneven usage across users; transfer to other products and languages requires further validation. Anthropomorphic stance is approximated by a single settings-layer naming act, which is neither a disposition scale nor a measure of discourse-level person reference. Treating every non-default name as anthropomorphic also collapses a real distinction: a person-name such as \emph{Ellie} and a functional label such as ``assistant~2'' both count as renaming, which makes the anthropomorphic group heterogeneous and, if anything, biases our comparisons toward the null.

All log-based results are associations. Naming almost always occurs at first contact, and we did not randomize the assistant name, so we cannot separate ``this kind of user would rename anyway'' from ``renaming changed how they talk''; identity tags are written from later conversation and likewise cannot be reversed into causes of naming. The pairwise judgments in RQ4 provide preference evidence, not causal evidence about long-term retention or task completion.

Two of our results are limited in opposite ways, and the difference is worth keeping apart. Several correlations in RQ3 are small enough to constrain little on their own (the identity tags run from $+0.07$ to $+0.27$), so they support a judgment about who is more likely to rename and not a design that acts on any one tag. The confiding gap has the reverse profile: it is the largest difference we observe, at more than twice the activation rate, but it is one latent among 128 and does not clear the Benjamini--Hochberg threshold, so we cannot exclude that a gap this size would turn up somewhere in a codebook of that width by chance. A confiding lexicon built independently of the autoencoder, or a replication on a second dictionary, would settle it; neither is available from the present logs. Latent short names were produced by a model at temperature 0 and cross-checked by a second prompt; we treat them as descriptive labels for reading the results and do not report a formal inter-coder agreement statistic. This departs from the practice of Purington et al., who report Cohen's $\kappa$ for their hand-coded personification measure~\cite{purington2017}, so any argument that depends on the exact wording of a latent name should be treated with corresponding caution. Sparse codebooks also have known failure modes in feature absorption and splitting~\cite{chanin2025absorption,bussmann2025matryoshka}, so an individual latent need not correspond to a single, monosemantic behavior.

The judgment sample (129 users) is smaller than the registered population (257), and heavy users contribute more judgments, so we report medians and interquartile ranges rather than one judgment per user; the reported percentages are therefore weighted toward frequent users, and we do not fit a user-level random-effects model. The design-effect bound in \S\ref{sec:rq43} brackets what that costs, but a model that estimated the intra-user correlation from the judgments themselves would place the tests more precisely than a pair of bounds. The left-hand reply is also a written specification of the previous register rather than a replay of the prompt that was actually deployed before the change; the two are meant to be equivalent on opening turns, and we did not run a separate arm to confirm that they are. Because the two branches are identical on tasking openings, a judgment generated on a tasking utterance compares two samples of one policy, and ``about the same'' is the expected answer there; the Q3 majority therefore reflects the mix of opening types in the judgment pool as well as the policy's effect on relational openings. Left and right positions were fixed in the evaluation material, so a position preference cannot be separated from a content preference; a production deployment should randomize them. Both groups saw the same fixed layout, however, so a position preference cannot account for the differences between the groups, which is what the Q1, Q2, and Q4 results rest on; the Q3 reading rests on the ``about the same'' share, which is not directional either. The evaluation also measures stated preference at the moment of the turn, not retention, task success, or any effect that would only appear over weeks. This gap matters most for the relational branch: recent work triangulating quasi-experiments with user accounts finds that companionship-oriented use of AI has measurable effects on wellbeing~\cite{aicompanion2026}, which a single-turn preference judgment cannot see. Finally, the sparse-autoencoder dictionary is frozen at a single seed~\cite{paulo2026sae}, and latents need not align across seeds, so the specific latent indices in this paper are properties of one trained dictionary rather than of the data alone.

\section{Conclusion}
\label{sec:conc}

Using three months of real usage from a desktop assistant, this paper examined the relationship between anthropomorphic stance and user behavioral patterns, and revised the reply policy accordingly. Naming in the settings layer splits in two: 45.9\% of users adopt an anthropomorphic name and the rest keep the default. The two groups' behavioral patterns over a shared codebook differ systematically ($\rho=+0.249$; permutation $p=.0419$). Teacher and tutoring identities covary positively with anthropomorphic naming and an introverted task-only orientation negatively, with employee and professional-exam preparation tags pointing the same way but not clearing that block's multiplicity correction; users who rename greet more, use fewer imperatives, and hold more conversation-only turns, and emotional confiding is the largest gap in the codebook. We therefore fork the reply by stance and by the current opening type: greet and receive affect on relational openings, execute directly on tasking openings. Across 1{,}744 in-product pairwise judgments from 129 users, those who renamed chose the forked reply far more often than default-name users did on being treated as a person and on having affect received, and less robustly on being willing to continue, while on the interference item neither group judged it worse than the register it replaced. The result shows that the behavioral difference is perceptible to users, that a policy conditioned on it is preferred by the group it targets, and that a choice users have already made can serve as a lightweight, observable conditioning variable.

\begin{acks}
A version of this paper is under review for CHI 2027.
\end{acks}

\bibliographystyle{ACM-Reference-Format}
\bibliography{refs}

\end{document}